\documentclass[seceq]{ptptex}
\usepackage{wrapft}

\newcommand{\pdi}[2]{\frac{\partial #1}{\partial #2}}

\newcommand{\sech}[0]{{\rm sech}}

\newcommand{\integ}[1]{\int d#1 \hspace*{1mm}}

\usepackage{amssymb}
\usepackage[dvips]{graphicx}
\usepackage{booktabs}
\usepackage{tabularx}
\usepackage{graphicx,color} 
\usepackage{wrapft}

\notypesetlogo
\title{Static Soliton at Nonequilibrium Steady State}

\author{
Shigeru \textsc{Ajisaka}\footnote{Email: g00k0056@suou.waseda.jp},
Shuichi \textsc{Tasaki},
and Ichiro \textsc{Terasaki}
}

\inst{
Advanced Institute for Complex Systems
and 
Department of Applied Physics,\\
School
of Science and Engineering,
Waseda University,\\
Tokyo 169-8555,
Japan
}

\abst{

Nonequlibrium phase transition of an open Takayama-Lin Liu-Maki chain coupled with two reservoirs is investigated.\
We will show that solitons connecting two uniform phases are possible, and the amplitude of solitons obeys the same self-consistent equation as that of uniform phases.
}

\begin{document}
\maketitle

\section{Introduction}
The study of nonequilibrium transport of electrons is essential to understanding the underlying physics of nano devices.
Because of the development of nano devices, theoretical description of quantum transport may lead to experimentation to test the validity of quantum theories.
As such, the pure science of quantum transport holds significant value in contemporary mesoscopic physics.

A common approach to the study of quantum transport in low-dimensional systems is putting systems with infinitely extended reservoirs into contact with one another.
However, an analytical approach to such systems is usually difficult, and the majority of works are performed numerically.
Nevertheless, recent progress in analytical frameworks allows us to study several systems at nonequilibriumn steady state (NESS).
The algebraic approach is promising for the study of NESS in such systems\cite{LecMath1880,Rue1,JP1}, and NESS is studied in a small number of cases, e.g., 
suppression of Fano-Kondo plateau in Aharonov-Bohm rings\cite{Takahasi}, nonlinear conductance of a solvable model of the Kondo effect\cite{Katsura}, long range correlations in the XY model\cite{HoAraki,Aschbacher}, and a new phase transition of the Takayama-Lin Liu-Maki (TLM) model\cite{PTP}.

Using a result of the algebraic approach\cite{Buttiker,PTP,Tasaki,Takahashi2} (Eq.~(\ref{ChNESS}) in \S 3), we studied  NESS of the TLM model\cite{PTP} as a representative example of a 1-D system exhibiting collective orders.
In that work, we discussed isotropic phases (i.e., uniformly dimerized chains) and the suppression of order under the application of current and bias voltage.
Also, we found orders to be multi-valued functions of bias voltage and single-valued functions of current.
Since the continuity equation implies that current is spatially uniform, 
the existence of a solution which connects two uniform domains with the equal amplitude is suggested.

Indeed, it is known that only solitons (amplitude kinks) connecting two domains with equal amplitudes are possible at equilibrium\cite{TLM}.

In this paper, we discuss such solitons at NESS.
As in the equilibrium case, the corresponding fermionic spectrum consists of continuum states and a midgap state at the center of the energy gap.

However, since our system is connected to reservoirs, reservoir fermions inside the energy gap should be considered.
We will show that in spite of the existence of the midgap state, reservoir fermions carrying the wave number inside the energy gap do not contribute to the gap equation, resulting in the same gap equation as the isotropic phases.

\section{Open TLM Model and the Equation of Motion}
The TLM model\cite{TLM} is a continuum limit of a tight-binding model (the SSH lattice) for polyacetylene proposed by Su, Schrieffer, and Heeger\cite{SSH1,SSH2} which describes charge density wave commensurate with the lattice. 

The TLM chain consists of two fermionic fields, $d(x)$ and $e(x)$, and the quantized local lattice distortion $\hat{\Delta}(x)$. 
Our Hamiltonian is given by (see appendix~A of ref.~\citen{PTP} for its derivation)
\begin{eqnarray*}
H_S &=&
\int_0^\ell dx
\Psi^\dag(x)
\left[-i \hbar v
\sigma_y\frac{\partial }{\partial x}+ 
\hat{\Delta} (x) \sigma_x
\right]
\Psi(x)
\nonumber \\
&&+
{1\over 2\pi \hbar v \lambda}\int_0^\ell dx 
\left[\hat{\Delta}(x)^2
+{1\over \omega_0^2}\hat{\Pi}(x)^2
\right]
\ ,
\end{eqnarray*}
where $\Psi(x)\equiv \left( d(x) , e(x) \right)^T$ 
is the two-component fermionic field, $\ell$ is the length of the system, $v$ is the Fermi velocity, $\sigma_x$ and $\sigma_z$ are the Pauli matrices, $\lambda$ is the dimensionless coupling constant, $\omega_0$ is the 
phonon frequency and $\hat{\Pi}(x)$ corresponds to the momentum conjugate to $\hat{\Delta}(x)$. 
Nonvanishing equal-time commutation relations among those operators are
\begin{eqnarray*}
\{d(x), d(y)^\dag\}=
\{e(x), e(y)^\dag\}=\delta(x-y) \ ,
\left[\hat{\Delta}(x),\hat{\Pi}(y)\right]=i \hbar^2 \pi \lambda v \omega_0^2 \delta(x-y) \ ,
\end{eqnarray*}
where $\{A, B\}=AB+BA$ and $[A,B]=AB-BA$. As the system is finite, fermionic waves are reflected back at the 
edges.
This effect is taken into account by a boundary condition:
\begin{equation*}
d(0)=0 \ , \ \ \ e(\ell)=0 \ .
\end{equation*}
The reservoirs are described by
\begin{equation*}
H_B=\int d{\boldsymbol k} \{ \hbar\omega_{kL} a_{{\boldsymbol k}L}^\dag a_{{\boldsymbol k}L} +
\hbar \omega_{kR} a_{{\boldsymbol k}R}^\dag a_{{\boldsymbol k}R}\} \ ,
\end{equation*}
where $a_{{\boldsymbol k}\nu}, (\nu=L,R)$ stands for the annihilation operators of reservoir fermions with momentum $\boldsymbol k$, $\hbar \omega_{k\nu}$ represents their energies, and their nonvanishing anticommutation relations are written $\{a_{{\boldsymbol k}\nu},a_{{\boldsymbol k}'\nu}^\dag\}
=\delta({\boldsymbol k}-{\boldsymbol k}')$.
The TLM chain-reservoir interaction is assumed to be
\begin{equation*}
V=\int d{\boldsymbol k} \ 
\bigg\{ \hbar v_{\boldsymbol k}
e^\dag(0) 
a_{{\boldsymbol k}L} + \hbar
w_{\boldsymbol k}
d^\dag(\ell) 
a_{{\boldsymbol k}R} + (h.c.) \bigg\} \ ,
\end{equation*}
where $v_{\boldsymbol k}$ and $w_{\boldsymbol k}$ stand for the coupling matrix elements. 
Then, the Hamiltonian of the whole system is given by
\begin{equation}
H=H_S+V+H_B
\ .
\label{TotalH}
\end{equation}
From (\ref{TotalH}), the lattice distortion $\hat{\Delta}$ is found to obey the following equation of motion: 
\begin{eqnarray}
\frac{\partial^2 \hat{\Delta}(x,t)}{\partial t^2}&=&
\frac{\partial \hat{\Pi}(x,t)}{\partial t}
=-\omega_0^2
\left(\hat{\Delta}(x,t)+\pi \hbar v \lambda
\Psi^\dag(x,t) \sigma_x \Psi(x,t)
\right)\ \ .
\label{DeltaEq}
\end{eqnarray}

\section{NESS Mean-Field Approximation}
In this section, we discuss mean-field NESS of the TLM model.
NESS under the mean-field approximation $\langle \cdots\rangle_\infty$
can be characterized as a state satisfying Wick's theorem with respect to the {\it incoming} fields 
$\alpha_{{\boldsymbol k}\nu}$ of $a_{{\boldsymbol k}\nu}$ for the mean-field Hamiltonian
\begin{eqnarray*}
H_{\rm \small MF}&=&H_S^{\rm \small MF}+V+H_B
\ ,
\\
H_S^{\rm \small MF}&\equiv&
\int_0^\ell dx
\Psi^\dag(x)
\left[-i \hbar v
\sigma_y\frac{\partial }{\partial x}+ 
\Delta(x) \sigma_x
\right]
\Psi(x) \ .
\end{eqnarray*}
Namely, they are defined as the solution of
\begin{eqnarray*}
&&{1\over \hbar} [\alpha_{{\boldsymbol k}\nu}, H_{\rm \small MF} ]=\omega_{k\nu}\alpha_{{\boldsymbol k}\nu}
\ ,\ \ \ 
e^{iH_{\rm \small MF}t/\hbar} a_{{\boldsymbol k}\nu} e^{-iH_{\rm \small MF}t/\hbar} \ e^{i\omega_{k\nu} t} 
\to \alpha_{{\boldsymbol k}\nu}
\
(t\to -\infty)
\label{InCom1}
\end{eqnarray*}
If asymptotic fields are complete, the incoming fields satisfy:
\begin{equation}\label{ChNESS}
\langle \alpha_{{\boldsymbol k}\nu}^\dag \alpha_{{\boldsymbol k}'\nu}
\rangle_\infty= f_\nu(\hbar\omega_{k \nu}) \delta({\boldsymbol k}-{\boldsymbol k}') \ ,
\end{equation}
where $\hbar\omega_{k \nu}$ is the single-particle energy of wave number $\boldsymbol k$,  
$ f_\nu(x)\equiv 1/(e^{(x-\mu_\nu)/T_\nu}+1)$ is the Fermi distribution function, $T_\nu$ 
is the initial temperature, and $\mu_\nu$ is the initial chemical potential of the reservoirs ($\nu=L,R$).

The self-consistent equation for the lattice distortion $\Delta(x)\equiv\langle \hat{\Delta}(x) \rangle_\infty$ is derived from the
equation of motion (\ref{DeltaEq}) for the lattice distortion and the time-independence of 
$\Delta(x)$.
Then, (\ref{DeltaEq}) leads to
\begin{eqnarray}\label{Self1}
0={-1\over \omega_0^2}\frac{\partial^2 \Delta(x)}{\partial t^2}&=&
\Delta(x)+\pi \hbar v \lambda
\langle
\Psi^\dag(x) \sigma_x \Psi(x)\rangle_\infty
\ .
\end{eqnarray}
Hereafter, we study a case where the lattice distortion is a soliton (amplitude kink) connecting two domains with the equal amplitude:
$\Delta(x)=\Delta_0\tanh \kappa_s (x-a)$.
Moreover, we restrict ourselves to the case of $\Delta_0=\hbar v\kappa_s$, which corresponds to cases in which creation energy of a soliton is minimum, and in which  only such solitons satisfy the self-consistent equation at equilibrium(see ref.~\citen{TLM}).
In this case, the self-consistent equation (\ref{Self1}) reads as
\begin{eqnarray}\label{Self2}
&&
{v\over\pi}
\int_{-\infty}^\infty d\omega 
\Biggl\{{\rm Im} \widetilde{\xi}_-(\omega)
{
h(x;\omega)^\dag \sigma_x
h(x;\omega) 
\over |\Lambda_-(\omega)|^2} f_L(\hbar\omega)
\nonumber\\
&&\mskip 140mu +{\rm Im} \widetilde{\eta}_-(\omega)
{
{\widetilde h}(x;\omega)^\dag
\sigma_x {\widetilde h}(x;\omega)
\over |\Lambda_-(\omega)|^2} f_R(\hbar\omega)
\Biggr\} 
=-
{\Delta(x)\over \pi \hbar v \lambda}
\ ,
\label{self}
\end{eqnarray}
where the integral is taken over by the reservoir fermions' energy, $\Lambda_-(\omega)$, $h(x;\omega)$ and ${\widetilde h}(x;\omega)$ are auxiliary functions defined by (the green function is given in appendix~\ref{app_green}, and relations between the auxiliary functions and the green function was shown in appendix~B of ref.~\citen{PTP})
\begin{eqnarray*}
D\Lambda_-(\omega)&=&
\Big[(\hbar v\kappa)+\frac{v\kappa}{\omega}\widetilde{\eta}_-
\left(\Delta(\ell)-\Delta(0)\right)
-\hbar v\kappa\widetilde{\xi}_-\widetilde{\eta}_-\Big]\cos\kappa\ell
\\&&\mskip 50 mu
+\Big[
\Delta(0)+\hbar\omega(\widetilde{\xi}_- +\widetilde{\eta}_-)
+\frac{-\Delta_0^2+\Delta(0) \Delta(\ell)}{\hbar\omega}
\widetilde{\eta}_-
+\Delta(\ell)\widetilde{\xi}_-\widetilde{\eta}_-
\Big]\sin\kappa\ell
\\
\begin{pmatrix} h_1(x) \\ h_2(x) \end{pmatrix}&=&
\frac{1}{vD}
	\begin{pmatrix}
	-\hbar v\kappa
	+\frac{v\kappa \left(\Delta(x)-\Delta(\ell)
	\right)}{\omega}\widetilde{\eta}_{+}
	\\
	 \hbar v\kappa \widetilde{\eta}_{+}
	\end{pmatrix}
\cos\kappa (x-\ell)
\\
&&\mskip 200 mu+
\frac{1}{vD}
	\begin{pmatrix}
	\Delta(x)
	+\frac{(\hbar v\kappa)^2+\Delta(x) \Delta(\ell)}
	{\hbar\omega}\widetilde{\eta}_{+}
	\\
	 \hbar\omega+ \Delta(\ell)\widetilde{\eta}_{+}
	\end{pmatrix}
\sin\kappa (x-\ell)
\\
\begin{pmatrix} \widetilde{h}_1(x) \\  \widetilde{h}_2(x) \end{pmatrix}&=&
\frac{1}{vD}
	\begin{pmatrix}
	-\frac{v\kappa}{\omega}\{ \Delta(x)-\Delta(0)\}
	+\hbar v\kappa\widetilde{\xi}_{+}
	\\
	 -\hbar v\kappa 
	\end{pmatrix}
\cos\kappa x
\\
&&\mskip 250 mu
-
\frac{1}{vD}
	\begin{pmatrix}
	\frac{\Delta(x) \Delta(0)+(\hbar v\kappa)^2}
	{\hbar\omega}
	+\Delta(x)\widetilde{\xi}_{+}
	\\
	\left(\Delta(0)+\hbar\omega\right) \widetilde{\xi}_{+}
	\end{pmatrix}
\sin\kappa x
\\
\widetilde{\xi}_\pm(\omega)&=& \frac{1}{v}\integ{{\bf k}'}\frac{|v_{{\bf k}'}|^2}{\omega-\omega_{k'L}\pm i0}
,\ \ \ 
\widetilde{\eta}_\pm(\omega)= \frac{1}{v}\integ{{\bf k}'}\frac{|w_{{\bf k}'}|^2}{\omega-\omega_{k'R}\pm i0}
\\
\kappa&=&\sqrt{(\hbar\omega)^2-\Delta^2}/(\hbar v),\ \ \ 
D(\omega)=\hbar\kappa v\cos \kappa \ell + \Delta(0)\sin \kappa \ell
\end{eqnarray*}

Equation (\ref{Self2}) is the self-consistent equation for the order parameter
$\Delta(x)$.
Note that we use a convention ${\rm Im}\widetilde{\xi}_-(\omega)=0$ (${\rm Im}\widetilde{\eta}_-(\omega)=0$) for $\omega$ 
outside the range of $\omega_{k\nu}$.
We further note that in accordance with $\Delta_0=\hbar v\kappa_s$, $\alpha_{k L}$ depends on position $x$ and $\alpha_{kR}$ does not depend on position in case of $\Delta_0=\hbar v\kappa_s$.

\section{Gap Equation for Solitons}

\subsection{Self-consistent Equation}
In this section, we show that the left-hand side of Eq.~(\ref{Self2}) is proportional to $\Delta(x)=\Delta_0 \tanh \kappa_s (x-a)$ for a long chain, and derive the self-consistent equation for solitons' amplitude $\Delta_0$.

Hereafter, we consider a case where the zero-bias chemical potential is 
located at 
the band center of the TLM chain. In other words, the chain is half-filled. This case corresponds
to $\hbar\omega_{k\nu}=(\hbar |{\boldsymbol k}|)^2/(2m_\nu)-\epsilon_{0\nu}$ ($\nu=L,R$) 
with $m_{\nu}$ and $\epsilon_{0\nu}$ being, respectively, the effective mass and the 
zero-bias chemical potential measured from the band bottom of the reservoirs.
In order to prevent the increase of electrostatic energy, the chemical potentials 
of the reservoirs should be chosen so that $\mu_L=-\mu_R=-eV/2$ with $V$ signifying bias voltage, and $e$ the elementary charge (for detail, see appendix~C of ref.~\citen{PTP}).
As is well known\cite{TLM}, the energy cutoff $\hbar\omega_c$ is necessary for the TLM model and we assume that $\epsilon_{0\nu}-e|V|/2>\hbar\omega_c$.
The integration interval of left-hand side of Eq.~(\ref{Self2}) should be replaced by $(-\omega_c,\omega_c)$.
In contrast to the isotropic dimerization, integrant of $|\hbar\omega|<\Delta_0$ should be carefully treated due to the existence of the midgap state.
In the following two subsections, we evaluate left-hand side. of Eq.~(\ref{self}).

\subsection{Reservoir Fermions with Energy inside Energy Gap}
In this subsection, the left-hand side of Eq.~(\ref{self}) is evaluated in the energy range of $|\hbar\omega|<\Delta_0$, which corresponds to the energy gap of the system.
This appears exponentially small, as in the case of uniform phases.
However, there exists a midgap state at the center of the gap, and the exponentially large term with respect to $\ell$ in the denominator $D\Lambda_-$ becomes comparable to the exponentially small term with respect to $\ell$.
Therefore, we employ the new integrating variable $\epsilon y=\hbar\omega$
 with $\epsilon\equiv\exp(-\kappa_s\ell)$ to scale up the behavior near $\omega=0$.
Then, $D\Lambda_-$ is evaluated by
\begin{eqnarray*}
2D\Lambda_-
=
\frac{1}{y}
\left[
\hbar \widetilde{\xi}_- y^2
+
\left(2e^{\kappa_s (\ell-2a)} \Delta_0-2e^{-\kappa_s (\ell-2a)}\Delta_0\widetilde{\xi}_- \widetilde{\eta}_-\right)y
+\frac{4\Delta_0^2}
{\hbar}\widetilde{\eta}_-
\right]
+O(\epsilon)\ ,
\end{eqnarray*}
where the first three terms are exponentially large, and the last term is exponentially small in the original integrating variable $\omega$.
Since we are considering a long chain, we focus on the center of the chain.
For this purpose, we further introduce a new variable $x=\ell/2+\delta x$.
It is thus easy to prove that after rewriting the left-hand side of Eq.~(\ref{self}) with $y$ and 
$\delta x$, terms on the order of $\epsilon^{-1}$ and $\epsilon^0$ vanish.
As a result, we conclude that the contribution of reservoir fermions with energy inside the energy gap to the gap equation is exponentially small, and thus is negligible.
Note that this is in contrast to the result of polarons which only appear out-of-equilibrium due to the contribution of reservoir fermions carrying energy inside the gap to the gap equation.
In that case, reservoir fermions near the midgap contribute to the gap equation and induce new excitation to the spinless TLM model (this aspect will be discussed elsewhere).

\subsection{Reservoir Fermions Carrying Energy inside Continuum State}
In this subsection, the left-hand side of Eq.~(\ref{self}) is evaluated in the energy range of $|\hbar\omega|>\Delta_0$.
Since the dominant terms of the left-hand side of Eq.~(\ref{self}) do not change regardless of energy, and because we consider a case where the center of the soliton is far from the chain end, i.e., $a=O(\ell),\ (\ell-a)=O(\ell)$, the lattice distortion at the chain ends, i.e., $\Delta(0)$ and $\Delta(l)$,  are approximated by $-\Delta$ and $\Delta$.
Then, e.g., the second term of the left-hand side of Eq.~(\ref{Self2}) reads
as
\begin{eqnarray}\label{Aux}
&&
\int_{\Delta_0/\hbar<|\omega|<\omega_c} d\omega\ 
v{\rm Im} \widetilde{\eta}_-(\omega)
{
{\widetilde h}(x;\omega)^\dag
\sigma_x {\widetilde h}(x;\omega)
\over |\Lambda_-(\omega)|^2} f_R(\hbar\omega) 
\nonumber\\
&&=\int_{\Delta_0/\hbar<|\omega|<\omega_c} d\omega 
\frac{{\rm Im} \widetilde{\eta}_-(\omega)\ f_R(\hbar\omega)}
{v|D(\omega)\Lambda_-(\omega)|^2}
\Big\{S_2(\omega)\sin 2\kappa x
+C_2(\omega)\cos 2\kappa x 
\nonumber\\
&&~~~~~~~~~~~~~~~~~~+\hbar \Delta(x)
\Big(\omega-{2\Delta_0 \over \hbar}{\rm Re}\widetilde{\xi}_-(\omega)
+\omega|\widetilde{\xi}_-(\omega)|^2\Big)\Big\}
,\ 
\end{eqnarray}
where $C_2(\omega)$ and $S_2(\omega)$ are defined by
\begin{eqnarray*}
C_2(\omega)&=&2
\left\{\Delta_0 \Delta(x)-(\hbar v\kappa)^2\right\}{\rm Re}\widetilde{\xi}_-(\omega)
-\hbar\omega\Delta(x)|\widetilde{\xi}_-(\omega)|^2
\nonumber\\
&&~~~~~~~~~~~+\frac{2 \Delta_0 (\hbar v\kappa)^2+\Delta(x)\left\{(\hbar v\kappa)^2-\Delta_0^2\right\}}{\hbar\omega}
\nonumber\\
S_2(\omega)&=&2(\hbar v\kappa)\left(\Delta(x)+\Delta_0\right){\rm Re}\widetilde{\xi}_-(\omega)
-\hbar\omega(\hbar v\kappa)|\widetilde{\xi}_-(\omega)|^2
\nonumber\\
&&~~~~~~~~~~~+(\hbar v\kappa)\frac{(\hbar v\kappa)^2-\Delta_0^2-2\Delta_0~\Delta(x)}{\hbar\omega}
\ .
\end{eqnarray*}

Moreover, by applying the Riemann-Lebesgue lemma, (\ref{Aux}) is found to be
\begin{eqnarray*}
&&
\int_{\Delta_0/\hbar<|\omega|<\omega_c} d\omega\ 
v~{\rm Im} \widetilde{\eta}_-(\omega)
{
{\widetilde h}(x;\omega)^\dag
\sigma_x {\widetilde h}(x;\omega)
\over |\Lambda_-(\omega)|^2} f_R(\hbar\omega) 
\nonumber\\
&&=\hbar \Delta(x)\int_{\Delta_0/\hbar<|\omega|<\omega_c} d\omega 
{{\rm Im} \widetilde{\eta}_-(\omega)\over v\zeta_0(\omega)}
f_R(\hbar\omega)
\Big(\omega-{2\Delta_0\over \hbar}{\rm Re}\widetilde{\xi}_-(\omega)
+\omega|\widetilde{\xi}_-(\omega)|^2\Big)
\ ,
\end{eqnarray*}
where $1/\zeta_0(\omega)$ corresponds to a dominant term of the Fourier coefficient of 
$1/|D\Lambda_-(\omega)|^2$, and is given by
\begin{eqnarray*}
\zeta_0(\omega) &=& 
\hbar^2 v\kappa 
\Bigg|
{\rm Im}\widetilde{\xi}_-(\omega)
\left\{ 
\omega+
\frac{2\Delta_0}{\hbar}
{\rm Re}\widetilde{\eta}_-(\omega) 
+\omega\big|\widetilde{\eta}_-(\omega)\big|^2
\right\}
\\
&&\mskip 70 mu + 
{\rm Im}\widetilde{\eta}_-(\omega)
\left\{ 
\omega-
\frac{2\Delta_0}{\hbar}{\rm Re}\widetilde{\xi}_-(\omega) 
+\omega\big|\widetilde{\xi}_-(\omega)\big|^2
\right\}\Bigg|\ .
\end{eqnarray*}
By a similar argument, when the TLM chain is long enough, 
the first term of the left-hand side of Eq.~(\ref{Self2}) is proportional to 
$\Delta(x)$, and Eq.~(\ref{Self2}) leads to
\begin{eqnarray}\label{SelfUni1a}
&& \Delta_0=0 \quad {\rm or}
\\
&&
{-1\over \lambda}=
S(\Delta_0,V,T_L,T_R)
\ ,
\label{SelfUni1b}
\end{eqnarray}
where 
\begin{eqnarray*}
&&
{S(\Delta_0,V,T_L,T_R)\over \hbar^2} \equiv
\int\limits_{{|\Delta_0|\over\hbar}<|\omega|<\omega_c} d\omega 
\Biggl\{
{{\rm Im} \widetilde{\xi}_-(\omega)\over \zeta_0(\omega)}
\Big(\omega+{2\Delta_0\over \hbar}{\rm Re}\widetilde{\eta}_-(\omega)
+\omega|\widetilde{\eta}_-(\omega)|^2\Big)
f_L(\hbar\omega)
\nonumber\\
&&\mskip 140mu 
+{{\rm Im} \widetilde{\eta}_-(\omega)\over \zeta_0(\omega)}
\Big(\omega-{2\Delta_0\over \hbar}{\rm Re}\widetilde{\xi}_-(\omega)
+\omega|\widetilde{\xi}_-(\omega)|^2\Big)
f_R(\hbar\omega)
\Biggr\} 
\ .
\end{eqnarray*}
When the TLM chain couples with two reservoirs at temperature $T=T_L=T_R$, an assumption $\widetilde{\eta}_-(\omega)^*=-\widetilde{\xi}_-(\omega)$ simplifies
the self-consistent equation~(\ref{SelfUni1b}):
\begin{eqnarray}
{1\over \lambda}=\int^{\hbar\omega_c}_{|\Delta_0|}
\frac{d \epsilon}{\sqrt{\epsilon^2-\Delta_0^2}}\
\frac{ \sinh(\epsilon/T)}
{\cosh(\frac{eV}{2T})+\cosh(\epsilon/T)}=-S(\Delta_0,V,T)
\label{SelfUniNum}
\end{eqnarray}
We would like to remark that the amplitude of solitons $\Delta_0$ and isotropic phases obey the same self-consistent equation (Eq.~(\ref{SelfUniNum})), though another assumption $\widetilde{\eta}_-(\omega)=\widetilde{\xi}_-(\omega)$ is required for isotropic phases (this difference is induced by the phase shifts of the solitons).

\subsection{Fermionic Current}
The fermionic current at $x$ in the TLM chain is given by
\begin{equation*}
\hat{J}(x)=-ev \Psi^\dag(x)\sigma_y\Psi(x) \ .
\end{equation*}
Since current is carried by fermions with energies near the fermi level, we approximate $\widetilde{\xi}_-(\omega),\ \widetilde{\eta}_-(\omega)$ by $\widetilde{\xi}_-(\omega)=\widetilde{\eta}_-(\omega)=i \Gamma/v$.
By applying the Riemann-Lebesgue lemma, its NESS average is reduced to
\begin{eqnarray}
J
=\frac{G_0}{e}
\int_{|\Delta_0|<|\epsilon|<\hbar\omega_c} d\epsilon 
\frac{\sqrt{\epsilon^2- \Delta_0^2}}
{|\epsilon|}
\left[f_R(\epsilon)-f_L(\epsilon)\right]
\ ,
\label{current}
\end{eqnarray}
where $G_0={e^2v\Gamma}/\{\pi\hbar(v^2+\Gamma^2)\}$ is 
the conductance in the normal phase.
We would like to note that the effect of fermions carrying energy inside the energy gap is exponentially small with respect to the chain size $\ell$, and is negligible.

\section{Conclusions}
We have studied nonequilibrium Peierls transition in the Takayama-Lin Liu-Maki chain connected to two reservoirs at different 
chemical potentials (their difference corresponds to bias voltage) by combining a mean-field approximation and the formula 
(\ref{ChNESS}), which is an outcome of the algebraic field approach to nonequilibrium statistical mechanics. 
We show that amplitude of solitons and isotropic lattice distortion obey the same self-consistent equation~(\ref{SelfUniNum}) after certain approximations; however, solitons require another condition  of tunneling between systems and reservoirs because of the effect of phase shifts.
We only consider solitons whose creation energy is minimum ($\Delta_0=\hbar v\kappa_s$).
However, not only such solitons are essential at low temperatures but also they are known as the only static solitons allowed at equilibrium.
At low temperatures, solitons' widths increase as a function of bias voltage $V$ and temperature $T$ (an outcome of a decrease of solitons' amplitude at low temperatures as discussed in ref.~\citen{PTP}).
Detailed behavior of the self-consistent equation (\ref{SelfUniNum}) and current between two reservoirs (\ref{current}) were particularly studied in ref.~\citen{PTP}.
Since the amplitude of solitons follows the same self-consistent equation as the isotropic lattice distortion, the stabilities studied in ref.~\citen{PTP} can be considered to be the same as the stabilities of the amplitude degree of freedom of the solitons.

Solitons' amplitudes or widths ($\Delta(x)=\hbar v \kappa_s$) characterize fermionic transport because current between reservoirs only depends on their amplitude and width.
To be more precise, higher bias voltage $V$ and/or higher temperature $T$ imply
smaller amplitude (larger width soliton) 
and lower current at low temperatures.
Detailed relationships between temperature, bias voltage and solitons' amplitude can be understood by replacing isotropic lattice distortion with solitons' amplitude, as in the previous paper.
The results of the isotropic lattice distortion, i.e., amplitude of the solitons, are roughly summarized as follows.
\begin{itemize}
\item{at constant bias voltage}\\
The phase transition between ordered and normal phases could be first or second order depending on the parameters.
In some parameter regions, the voltage-current characteristics are S-shaped (namely, there is negative differential conductivity). Negative differential conductivity appears when temperature is lower than a certain  threshold value.
\item{at constant current}\\
All the non-trivial solutions of the self-consistent equation are stable and the phase transition between the ordered and normal phases is always second order.
\end{itemize}

\section*{Acknowledgements}
The authors thank T. Prosen, G. Casati, G. Benenti,
Y. Matsunaga, Baowen Li, Bambi Hu, K. Nakamura, A. Sugita, and N. Weissburg for fruitful discussions.
This work is partially supported by a Grant-in-Aid for Scientific
Research (Nos. 17340114, 16076213, and 17540365) from the Japan Society of 
the Promotion of Science,
for the ``Academic Frontier'' Project at Waseda University 
and the 21st Century COE Program at Waseda University ``Holistic
Research and Education Center for Physics of Self-Organization Systems''
both from the Ministry of Education, Culture, Sports, Science and
Technology of Japan.

\appendix
\section{Derivation of a continuous model \label{app_green}}
In this appendix, we explicitly show the Green function in the soliton case.
For the TLM model, the Green function obeys
\begin{eqnarray*}
D_+ g_{+\sigma}(x,y,\omega)&=\omega g_{-\sigma}(x,y,\omega)-\delta_{-\sigma}\delta(x-y)
\\
D_-g_{-\sigma}(x,y,\omega)&=\omega g_{+\sigma}(x,y,\omega)-\delta_{+\sigma}\delta(x-y)\ ,
\end{eqnarray*}
where $D_{\pm}$ is defined by $D_{\pm}\equiv \pm v\pdi{}{x}+\Delta(x)/\hbar$. Then, $g_{-\sigma}(x,y:\omega)$ follows
\begin{eqnarray*}
D_+ D_- g_{-\sigma}(x,y:\omega)&=&\Bigg[-v^2\pdi{^2}{x^2}+
\frac{v}{\hbar}\frac{d \Delta(x)}{dx}+\frac{\Delta(x)^2}{\hbar^2} \Bigg] g_{-\sigma}(x,y:\omega)\ ,
\end{eqnarray*}
at $x\neq y$. For the soliton $\Delta(x)=\Delta_0 \tanh \kappa_s(x-a)$, the operator $D_+ D_-$ is easily caluculated;
\begin{eqnarray*}
D_+ D_-=
-v^2\pdi{^2}{x^2}+
\frac{\Delta_0}{\hbar} v\kappa_s\ \sech^2\kappa_s(x-a)+
\frac{\Delta_0^2}{\hbar^2}\tanh\kappa_s(x-a)
\label{green eq}
\end{eqnarray*}
In general, the solutions of (\ref{green eq}) are expressed in terms of the hypergeometric function.
In this paper, we restrict ourselves to the case of 
$\Delta_0=\hbar v \kappa_s$. As such, the green function is given by
\begin{eqnarray*}
&&g_{++}(x,y;\omega)= 
\begin{cases} \displaystyle
\frac{ 
\big\{(\hbar v\kappa)^2+\Delta(0)\Delta(x)\big\}\sin \kappa(x)+\hbar v\kappa\big\{\Delta(x)-\Delta(0)\big\}\cos \kappa(x)}
{(\hbar v)^2 \kappa \omega D(\omega)} 
\\ \mskip 150 mu\times \Big[\Delta(y)\sin \kappa(y-l)-\hbar v\kappa\cos\kappa (y-l)\Big]
&(x<y) \cr\cr
\displaystyle
\frac{ 
\Delta(x)\sin\kappa (x-\ell)-\hbar v\kappa\cos\kappa (x-\ell)
}
{(\hbar v)^2 \kappa \omega D(\omega)} 
\\ \mskip 150 mu\times \Big[
\big\{(\hbar\kappa v)^2+\Delta(0)\Delta(y)\big\}
\sin\kappa y+
\hbar v \kappa\big\{ \Delta(y)-\Delta(0) \big\}
\Big]
&(x>y)\cr
\end{cases}
\\ 
\nonumber \\
&&g_{--}(x,y;\omega)= 
\begin{cases} \displaystyle
{\omega 
\big(\hbar v \kappa\cos \kappa x+\Delta(0)\sin \kappa x\big)\sin \kappa (y-\ell)
\over v^2 \kappa D(\omega)} &(x<y)\cr\cr
\displaystyle
{\omega 
\sin \kappa (x-\ell)
\big(\hbar v\kappa\cos \kappa y+\Delta(0)\sin \kappa y\big)
\over v^2\kappa D(\omega)} &(x>y)\cr
\end{cases}
\\ 
\nonumber\\
&&g_{+-}(x,y;\omega)=g_{-+}(y,x;\omega)\nonumber\\
&&~=\begin{cases} \displaystyle
{\Big[\big\{ (\hbar v\kappa)^2+\Delta(x) \Delta(0)\big\}\sin\kappa x+\hbar v\kappa \big\{ \Delta(x)-\Delta(0)\big\}
\cos\kappa x\Big]
\sin\kappa(y-\ell)
\over 
\hbar v^2\kappa D(\omega)} 
&(x<y)\cr\cr
\displaystyle
\frac{\big(\Delta(x)\sin \kappa(x-\ell)-\hbar v\kappa\cos \kappa (x-\ell)\big)
\big(\Delta(0)\sin \kappa y+\hbar v\kappa\cos \kappa y\big)}
{\hbar v^2\kappa D(\omega)} 
&(x>y)\cr
\end{cases}
\end{eqnarray*}
where $\kappa=\sqrt{(\hbar\omega)^2-\Delta_0^2}/(\hbar v)$ and 
$D(\omega)=\hbar v\kappa\cos \kappa \ell+\Delta(0)\sin \kappa \ell$.

\bibliographystyle{acm}
\bibliography{soliton2}


%

\end{document}